\documentclass[journal]{IEEEtran}

\usepackage[utf8]{inputenc}
\usepackage{blindtext}
\usepackage{graphicx}
\usepackage{color}
\usepackage{hyperref}
\usepackage{filecontents}

\usepackage[noadjust, nospace]{cite}

\usepackage[cmex10]{amsmath}
\usepackage{bbm}
\usepackage{amssymb}
\usepackage{algorithmic}
\usepackage{array}
\usepackage{url}
\makeatletter
\@ifclasslater{IEEEtran}{2012/12/26}
 {\newcommand{\killproofname}{\unskip\nopunct}}
 {\newcommand{\killproofname}[1]{\unskip\aftergroup\ignorespaces\ignorespaces}}
\makeatother

\usepackage{graphicx}
\usepackage{subcaption}
\usepackage{subfiles}
\usepackage{amsfonts}
\usepackage{todonotes}
\usepackage{textcomp}
\usepackage{gensymb}
\usepackage{cleveref}

\usepackage{svg, import}

\usepackage{amsthm,amsmath}
\usepackage{bm}
\let\labelindent\relax
\usepackage{enumitem}

\newtheorem{thm}{Theorem}
\newtheorem{lemma}{Lemma}

\newtheorem{corollary}[thm]{Corollary}

\theoremstyle{definition}
\newtheorem{definition}{Definition}

\makeatletter
\@addtoreset{proofpart}{thm}
\makeatother

\usepackage{algorithm}

\def\dvbar#1{\Vert#1\Vert}

\def\norm#1{\dvbar{#1}}

\DeclareMathOperator{\sinc}{sinc}

\newcommand{\ind}[1]{^{\left(#1\right)}}
\newcommand{\vect}[1]{\mathbf{#1}}
\newcommand{\mat}[1]{\mathbf{#1}}
\newcommand{\spikes}[1]{\left\lbrace t_\ell\ind{#1}, \ell = 1\cdots n^{(#1)}_{\mathrm{spikes}} \right\rbrace}

\newcommand{\coeffs}[1]{\vect{C}(\vect{#1})}

\newcommand{\bias}{\beta}
\newcommand{\biasi}[1]{\beta\ind{i}}

\newcounter{assumptCounter}

\newenvironment{assumptions}{
    \begin{enumerate}[label = (A\arabic*)]
    \setcounter{enumi}{\value{assumptCounter}}
    }
    {
    \setcounter{assumptCounter}{\value{enumi}}
    \end{enumerate}
    }

\newenvironment{subassumptions}{
    \begin{enumerate}[label = (A\theassumptCounter.\alph*)]
    \setcounter{assumptCounter}{\value{assumptCounter}+1}
    }
    {
    \end{enumerate}
    }

\usepackage{todonotes}
\usepackage{tikz}
\usetikzlibrary{math}
\graphicspath{{figures/}{../figures/}}

\begin{document}

\setcounter{assumptCounter}{0}
\title{Asynchrony Increases Efficiency: \\Time Encoding of Videos and Low-Rank Signals}

\def\AFF{Karen~Adam, Adam~Scholefield and Martin~Vetterli are with the School of Computer and Communication Sciences, Ecole Polytechnique F\'{e}d\'{e}rale de Lausanne (EPFL), CH-1015 Lausanne, Switzerland, email: firstname.lastname@epfl.ch.}

\def\FUND{This work was in part supported by the Swiss National Science Foundation grant number 200021\_181978/1, ``SESAM - Sensing and Sampling: Theory and Algorithms''.}
\author{
Karen~Adam,~\IEEEmembership{Student Member,~IEEE,}
Adam~Scholefield,~\IEEEmembership{Member,~IEEE,}
Martin~Vetterli,~\IEEEmembership{Fellow,~IEEE}
\thanks{\FUND}\thanks{\AFF}}

\maketitle
\begin{abstract}
In event-based sensing, many sensors independently and asynchronously emit events when there is a change in their input. Event-based sensing can present significant improvements in power efficiency when compared to traditional sampling, because (1) the output is a stream of events where the important information lies in the \textit{timing} of the events, and (2) the sensor can easily be controlled to output information only when interesting activity occurs at the input.

Moreover, event-based sampling can often provide better resolution than standard uniform sampling. Not only does this occur because individual event-based sensors have higher temporal resolution~\cite{rebecq2019high}, it also occurs because the asynchrony of events allows for less redundant and more informative encoding. We would like to explain how such curious results come about.

To do so, we use ideal time encoding machines as a proxy for event-based sensors. We explore time encoding of signals with low rank structure, and apply the resulting theory to video. We then see how the \textit{asynchronous} firing times of the time encoding machines allow for better reconstruction than in the standard sampling case, if we have a high spatial density of time encoding machines that fire less frequently.
\end{abstract}


\section{Introduction}

Many aspects of our lives are governed by routine and rythm: our work days, circadian rythms, breathing patterns, or even music. However, applying metronomic schedules might not necessarily be resource-efficient for all applications. 
We generally say ``hello'' \emph{when} we see someone we know rather than saying it at regular intervals.

Many engineered systems, such as traditional sampling devices, rely almost exclusively on clocked behavior.
These sampling schemes are powerful - they govern how we record music, take images, transfer information -  but they fail to adapt their activity to the varying complexity of the input.

This drawback leads to inefficiencies which are apparent when comparing the power consumption of human-engineered technologies to biological equivalents.



As demands for increased storage and processing coupled with smaller devices has brought efficiency into the spotlight, researchers have been turning to biology for inspiration.

Inspired by neurons, event-based sensing is growing in popularity~\cite{delbruck2010activity, liu2010neuromorphic, gallego2019event}. The output of such a sensor is a series of spikes which are characterized by their \emph{timing} rather than their amplitude, as is the case with traditional sampling~\cite{posch2014retinomorphic}. In addition, since spikes times are dependent on the input, the activity of the output is correlated to the activity of the input.

    \begin{figure*}[tb]
        \centering
\begingroup%
  \makeatletter%
  \providecommand\color[2][]{%
    \errmessage{(Inkscape) Color is used for the text in Inkscape, but the package 'color.sty' is not loaded}%
    \renewcommand\color[2][]{}%
  }%
  \providecommand\transparent[1]{%
    \errmessage{(Inkscape) Transparency is used (non-zero) for the text in Inkscape, but the package 'transparent.sty' is not loaded}%
    \renewcommand\transparent[1]{}%
  }%
  \providecommand\rotatebox[2]{#2}%
  \newcommand*\fsize{\dimexpr\f@size pt\relax}%
  \newcommand*\lineheight[1]{\fontsize{\fsize}{#1\fsize}\selectfont}%
  \ifx\svgwidth\undefined%
    \setlength{\unitlength}{405.11624179bp}%
    \ifx\svgscale\undefined%
      \relax%
    \else%
      \setlength{\unitlength}{\unitlength * \real{\svgscale}}%
    \fi%
  \else%
    \setlength{\unitlength}{\svgwidth}%
  \fi%
  \global\let\svgwidth\undefined%
  \global\let\svgscale\undefined%
  \makeatother%
  \begin{picture}(1,0.56580745)%
    \lineheight{1}%
    \setlength\tabcolsep{0pt}%
    \put(0,0){\includegraphics[width=\unitlength,page=1]{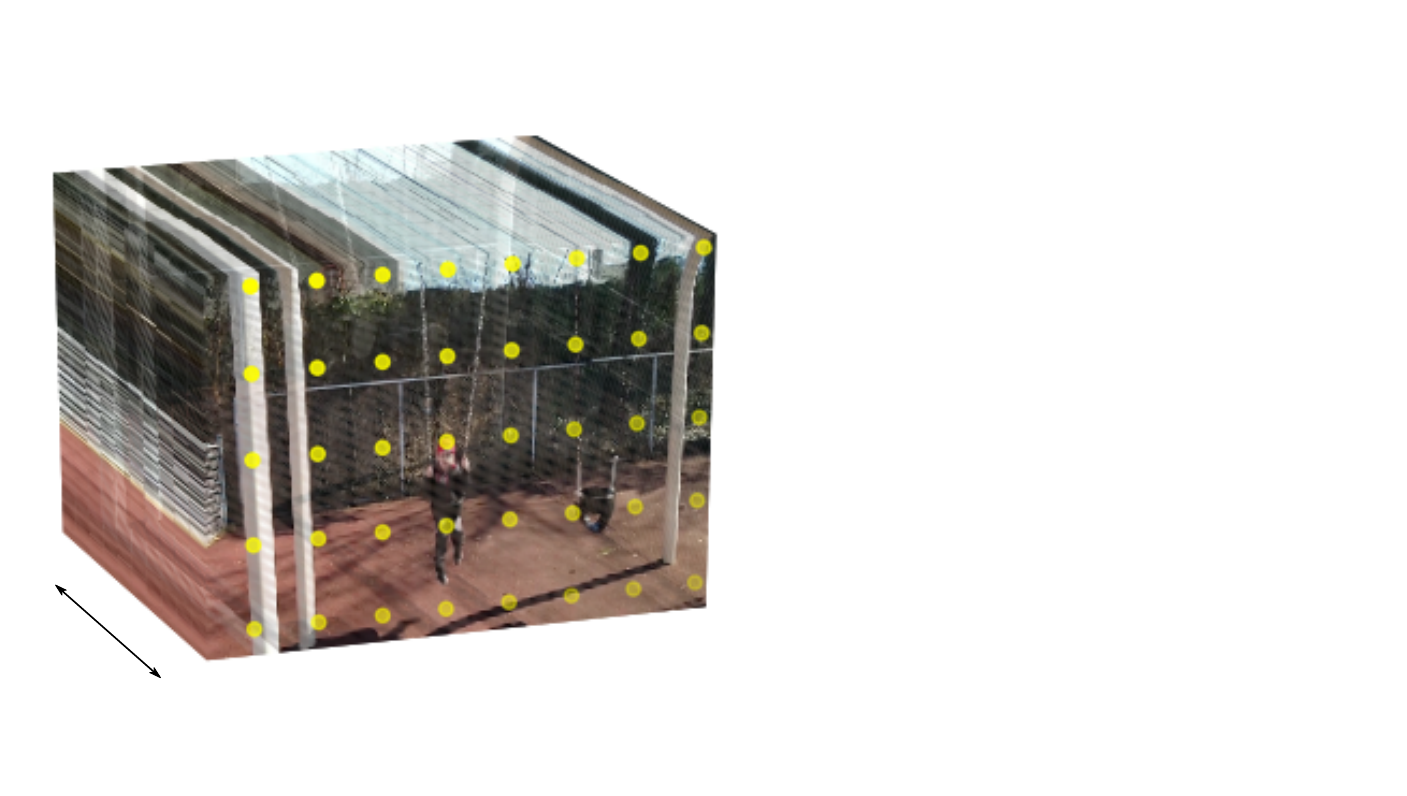}}%
    \put(0.02091409,0.11829427){\color[rgb]{0,0,0}\rotatebox{-37.210289}{\makebox(0,0)[lt]{\lineheight{1.25}\smash{\begin{tabular}[t]{l}Time\end{tabular}}}}}%
    \put(0,0){\includegraphics[width=\unitlength,page=2]{Figure1_try2_pdf.pdf}}%
    \put(0.81677981,0.00823734){\color[rgb]{0,0,0}\makebox(0,0)[lt]{\lineheight{1.25}\smash{\begin{tabular}[t]{l}Time\end{tabular}}}}%
    \put(0.81677981,0.41182572){\color[rgb]{0,0,0}\makebox(0,0)[lt]{\lineheight{1.25}\smash{\begin{tabular}[t]{l}Time\end{tabular}}}}%
    \put(0.81677981,0.27853084){\color[rgb]{0,0,0}\makebox(0,0)[lt]{\lineheight{1.25}\smash{\begin{tabular}[t]{l}Time\end{tabular}}}}%
    \put(0.81677981,0.14523545){\color[rgb]{0,0,0}\makebox(0,0)[lt]{\lineheight{1.25}\smash{\begin{tabular}[t]{l}Time\end{tabular}}}}%
    \put(0,0){\includegraphics[width=\unitlength,page=3]{Figure1_try2_pdf.pdf}}%
  \end{picture}%
\endgroup%

    \caption{Vision setup: we assume that we have an array of spiking devices, such as photoreceptors or TEMs, each of which is observing a scene at a particular location. The input to the receptor at this location is a time varying signal and the receptor will output a stream of spikes, the timing of which is dependent on the input.
    On the left, we show the projection of the scene which is being observed, with an overlay of event-based sensors shown in yellow. To its right, we zoom in to view the spiking output of some of the sensors.}
    \label{fig: Figure 1}
    \end{figure*}

While efforts have recently been invested to better understand event-based sensing and reconstruction of bandlimited or finite-rate-of-innovation signals~\cite{thao2020time,alexandru2019reconstructing,rudresh2020time},
comparisons between event-based sensing and standard sampling have mostly considered the timing-based output of event based sensing to be more of a pesky necessity that requires some work-around rather than a blessing in disguise. Actually, it is precisely this staggered asynchrony in the outputs that can allow for better resolution and more flexibility when using event-based sampling.


We show this in a series of steps. First, we review time encoding machines (TEMs), which are the ideal event-based sensor~\cite{lazar2004perfect,gontier2014sampling}. The input-output relationship of an integrate-and-fire TEM follow similar rules to that of nonuniform sampling~\cite{lazar2004timerefractory,feichtinger1994theory}, and thus results in interesting consequences when it comes to single-signal multi-channel time encoding~\cite{lazar2005multichannel,adam2019multichannel}.

The natural extension to multi-signal multi-channel time encoding then offers optimal sampling efficiency for low-dimensional signals with known structure. We find a Nyquist-like criterion on the number of spikes needed for reconstruction, requiring as many linearly independent constraints as degrees of freedom.

Using this formulation, the video recording problem with TEMs turns into a parametric estimation problem. 
We study the setup in Fig.~\ref{fig: Figure 1}: multiple TEMs are used to encode multiple locations in a scene and each TEM outputs a series of asynchronous spikes. We find that this setup offers interesting tradeoffs in terms of time and space resolution: for once, increasing space resolution can actually \emph{increase} time resolution as well, precisely thanks to our blessing in disguise---the staggered, asynchronous outputs of the TEMs.

As a result, time encoding or event-based sensing encourages increasing the number of sensors rather than the spiking rate of sensors as a means to improve resolution in both time and space. This, in turn, means that (1) there are fewer hardware requirements on the sensors themselves, and (2) sensors can integrate information over more time, thus avoiding issues with low photon count that occur at higher shutter speeds with standard cameras.

This particular result does not usually hold in the equivalent scenario in standard frame-based video recording. There, the resolution in time and space are almost independent from one another. The former depends on the frame-rate of the camera and the latter depends on the sampling pattern of the camera, which of course includes the number of pixels used for sampling. If we increase the number of pixels used to record a frame, this would not improve the resolution in time, because all pixels of a frame are taken at the same \textit{moment} in time. To fix this, different pixels would need to record frames at \emph{different} times. This is possible but renders things more complicated: reconstruction would either require that time shifts between pixel clocks be known, or it becomes a difficult problem that has no uniqueness guarantees. In contrast, when using time encoding, spike times are---by design---almost surely different and this difference comes at no extra cost.

\section{Background}
\label{sec: background}

As first presented in~\cite{lazar2003time}, time encoding machines (TEMs) encode inputs using \emph{times} that are dependent on the input itself. TEMs can therefore be used to model neurons or sensory receptors such as photoreceptors. In fact, the simpler neuron models often encode their input currents using action potentials with fixed amplitude and varying timing, where the timing holds the information about the input~\cite{burkitt2006review1}.

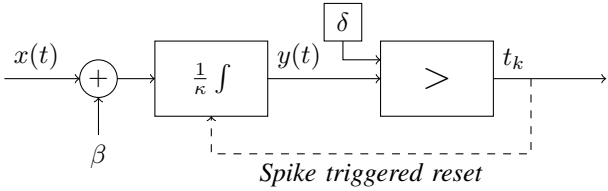
\begin{figure}[tb]
	\begin{minipage}[b]{0.85\linewidth}
		\centering
        \begin{tikzpicture}[scale = 1]

    \tikzmath{
        \inputstartx = 0;
        \inputstarty = 0.4;
        \inputlen = 1;
    }
    \draw[->] (\inputstartx, \inputstarty) node[anchor = south west] {$x(t)$} -- (\inputlen, \inputstarty);

    \tikzmath{
        \circleradius = 0.25;
        \circlecenterx = \inputstartx+\inputlen+\circleradius;
        \circlecentery = \inputstarty;
    }
    \draw (\circlecenterx, \circlecentery) circle (\circleradius) node {$+$};

    \tikzmath{
        \biaslinelen = 0.5;
        \biaslinex = \circlecenterx;
        \biasliney = \circlecentery -  \circleradius - \biaslinelen;
    }
    \draw[->] (\biaslinex, \biasliney) node[below] {$\bias$} --  (\biaslinex, \biasliney+ \biaslinelen);

    \tikzmath{
        \tointlinex = \circlecenterx + \circleradius;
        \tointliney = \circlecentery;
        \tointlinelen = 0.5;
    }
    \draw[->] (\tointlinex, \tointliney) -- (\tointlinex + \tointlinelen, \tointliney);

    \tikzmath{
        \integboxheight = 1;
        \integboxwidth = 1.5;
        \integboxx = \tointlinex + \tointlinelen;
        \integboxy = \tointliney - \integboxheight/2;
    }
    
    \draw (\integboxx, \integboxy) rectangle (\integboxx+\integboxwidth, \integboxy+\integboxheight) node[midway] {$\frac{1}{\kappa}\int$};

    \tikzmath{
        \tocomplinex = \integboxx + \integboxwidth;
        \tocompliney = \tointliney;
        \tocomplinelen = 1.5;
    }
    \draw[->] (\tocomplinex, \tocompliney) node[anchor = south west] {$y(t)$} -- (\tocomplinex + \tocomplinelen, \tocompliney);
    
    \tikzmath{
        \compboxheight = 1;
        \compboxwidth = 1.5;
        \compboxx = \tocomplinex + \tocomplinelen;
        \compboxy = \tocompliney - \compboxheight/2;
    }
    
    \draw (\compboxx, \compboxy) rectangle (\compboxx+\compboxwidth, \compboxy+\compboxheight) node[midway] {\Large $>$};
    
    \tikzmath{
        \deltalinkx = \compboxx;
        \deltalinky = \compboxy + 0.75\compboxheight;
        \deltalinkwidth = 0.5;
        \deltalinkheight = 0.25;
    }
    \draw[->] (\deltalinkx - \deltalinkwidth, \deltalinky + \deltalinkheight) -- (\deltalinkx- \deltalinkwidth, \deltalinky) -- (\deltalinkx, \deltalinky);
    
    \tikzmath{
        \deltaboxwidth = 0.5;
        \deltaboxheight = 0.5;
        \deltaboxx = \deltalinkx - \deltalinkwidth - \deltaboxwidth/2;
        \deltaboxy = \deltalinky + \deltalinkheight ;
    }
    
    \draw (\deltaboxx, \deltaboxy) rectangle  node {$\delta$} (\deltaboxx + \deltaboxwidth, \deltaboxy+ \deltaboxheight);
    
    \tikzmath{
        \tooutlinex = \compboxx + \compboxwidth;
        \tooutliney = \tocompliney;
        \tooutlinelen = 1.5;
    }
    \draw[->] (\tooutlinex, \tooutliney) node[anchor = south west] {$t_k$} -- (\tooutlinex + \tooutlinelen, \tooutliney);
    
    \tikzmath{
        \feedbacklinex = \tooutlinex +0.5;
        \feedbackliney = \tooutliney;
        \feedbacklineheight = 1;
        \feedbacklineendy = \integboxy;
        \feedbacklineendx = \integboxx + \integboxwidth/2;
    }
    \draw[dashed, ->] (\feedbacklinex, \feedbackliney) -- (\feedbacklinex , \feedbackliney- \feedbacklineheight) -- node[below] {\textit{Spike triggered reset}} (\feedbacklineendx , \feedbackliney - \feedbacklineheight) --    (\feedbacklineendx, \feedbacklineendy);
\end{tikzpicture}
	\end{minipage}
	\vspace{-2em}
	\caption{Circuit of a Time Encoding Machine, with input $x(t)$, threshold $\delta$, integrator constant $\kappa$ and bias $\bias$.}
	\label{fig:TEM circuit}
	\end{figure}

In this paper, we will consider one model for time encoding machines which resembles an integrate-and-fire neuron with no leak~\cite{lazar2004timerefractory, burkitt2006review1}. Such TEMs can provide perfect encodings of signals using one or many channels.
The circuit of a TEM is depicted in Fig.~\ref{fig:TEM circuit}.

\begin{definition}
    A \textit{time encoding machine} (TEM) with parameters $\kappa$, $\delta$, and $\bias$ takes an input signal $x(t)$, adds a bias $\bias$ to it and integrates the result, scaled by $1/\kappa$, until a threshold $\delta$ is reached. Once this threshold is reached, the time $t_k$ at which it is reached is recorded, the value of the integrator resets to $-\delta$ and the mechanism restarts. We say that the machine spikes at the integrator reset and call the recorded time $t_k$ a \textit{spike time}. 
\end{definition}

The first results on time encoding machines that resemble this model were, to the authors' knowledge, established by Lazar and T\'oth~\cite{lazar2004perfect}. 

The results operate under the following assumptions.
\begin{assumptions}
\item \label{ass: BL} The input signal $x(t)$ is bandlimited with bandwidth $\Omega$\label{enum: assumption: bandlimited}.
\item \label{ass: L2} The input signal $x(t)$ is in $\mathcal{L}^2(\mathbb{R})$\label{enum: assumption: L2}.
\item \label{ass: bounded} The input signal $x(t)$ is bounded by a constant $c$, $|x(t)|<c, \forall t \in \mathbb{R}$. \label{enum: assumption: bounded}
\end{assumptions}

Under these assumptions, the input $x(t)$ can be reconstructed from the emitted spike times if the parameters of the machine satisfy $\bias>c$ and the bandwidth satisfies 
\begin{equation}
\label{eq: single TEM condition on bandwidth}
\Omega < \frac{\pi(\bias-c)}{2\kappa\delta}.
\end{equation}

The reconstruction scheme and the proof of convergence are based on two key elements: 
\begin{enumerate}
\item the time encoding scheme is tightly related to the scheme of sampling averages, therefore the results developped for the reconstruction from averages can be used for time encoding and reconstruction~\cite{feichtinger1994theory, aldroubi2002non}, and
\item when performing time encoding, the maximal delay between two consecutive spike times  is dictated by the parameters of the machine:
\begin{equation}
t_{k+1}-t_k < \frac{2\kappa\delta}{\bias-c}.
\end{equation}
\end{enumerate}

Given these two observations,  and under Condition~\eqref{eq: single TEM condition on bandwidth}, the input signal can be perfectly determined by the spike times using algorithms based on alternating projections onto convex sets~\cite{lazar2004perfect,bauschke1996projection,thao2020time}.

Later, the theory was extended to multi-channel time encoding of a signal. On one hand, Lazar suggested a scheme for bandlimited signal sampling and reconstruction using many time encoding machines coupled with filter banks~\cite{lazar2005multichannel}. On the other hand, we suggested a scheme for sampling and reconstructing a bandlimited signal using many time encoding machines that are similar and have no pre-filters~\cite{adam2020sampling}. In the latter scenario, we showed that if one TEM can encode a signal with bandwidth $\Omega$, then $M$ TEMs can encode a signal with bandwidth $M\Omega$ assuming that the TEMs have \emph{unknown} non-zero shifts between their integrators $\alpha_1, \cdots, \alpha_M$~\cite{adam2020sampling}. In other words, $M$ time encoding machines with parameters $\kappa$, $\delta$, and $\bias$ can encode a signal which satisfies assumptions~\ref{enum: assumption: bandlimited},~\ref{enum: assumption: L2},~\ref{enum: assumption: bounded} if
\begin{equation}
\Omega < M\frac{\pi(\bias-c)}{2\kappa\delta}.
\end{equation}

As a result, instead of using one TEM with a certain spiking rate to encode a signal, one can now use many TEMs with lower spiking rates to encode the same signal. This is useful if time encoding machines or neurons have an upper limit on their spiking rate.

The result generalizes to multi-signal, multi-channel time encoding, as partly studied in~\cite{adam2020encoding} and as we will see in the next sections.


\section{Problem Setup}
\label{sec: Problem Setup}
For the remainder of this paper, we consider many time-varying signals $y\ind{i}(t), i = 1\cdots I$ that are correlated with each other. These signals are encoded using time encoding machines and we assume that each signal follows a parametric model which we know.

Correlated signals arise in many applications such as, among others, meteorological data, biomarkers in human patients, regional economic data, audio, and video. The latter example will be given particular attention later on.

    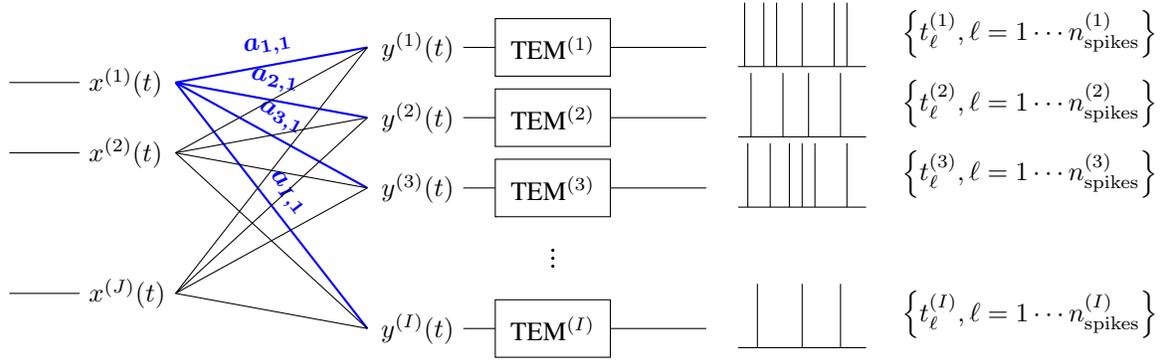
\begin{figure*}[tb]
        \centering
        \begin{center}
\begin{tikzpicture}[scale = 0.85]
\foreach \n in {0,1,2,3}
{
    
    \tikzmath{
        int \indn;
        \indn= \n +1;
        \startx = 1;
        \starty = 1 - \n*(0.9+0.2);
    }
    \ifnum \n = 3
        \tikzmath{
            \starty = 1 - 4*(0.9+0.2);
        }
    \fi
    \tikzmath{
        \endx = 1 + 1.8;
        \endy = \starty + 0.9;
        \linestartX = (\startx - 0.5);
        \lineY = (\starty+\endy)/2);
        \TEMLabelX = (\startx+\endx)/2;
    }
    \draw (\startx, \starty) rectangle (\endx, \endy);
    \ifnum \n < 3
        \draw (\TEMLabelX, \lineY) node {TEM$\ind{\indn}$};
    \else
        \draw (\TEMLabelX, \lineY) node {TEM$\ind{I}$};
    \fi
    \draw (\linestartX, \lineY) -- (\startx, \lineY) ;
    \draw (\startx+1.8, \lineY) -- (\startx+1.8+1.5, \lineY);
    \tikzmath{
        \startSpikex = \startx+1.8+1.5 + 0.5;
        \startSpikey = \lineY - 0.3;
    }
    \ifnum \n = 0
        \foreach \loc in {0.1,0.4,0.6,1,1.5, 1.7}
        {
            \draw (\startSpikex + \loc, \startSpikey) -- (\startSpikex + \loc, \startSpikey + 1);
        }
    \fi
    \ifnum \n = 1
        \foreach \loc in {0.2,0.7,1.1, 1.6}
        {
            \draw (\startSpikex + \loc, \startSpikey) -- (\startSpikex + \loc, \startSpikey + 1);
        }
    \fi
    \ifnum \n = 2
        \foreach \loc in {0.15,0.5,0.8,1,1.2,1.7}
        {
            \draw (\startSpikex + \loc, \startSpikey) -- (\startSpikex + \loc, \startSpikey + 1);
        }
    \fi
    \ifnum \n = 3
        \foreach \loc in {0.3,1, 1.6}
        {
            \draw (\startSpikex + \loc, \startSpikey) -- (\startSpikex + \loc, \startSpikey + 1);
        }
    \fi
    
    \ifnum \n <3
    \draw (\linestartX, \lineY) node[anchor = east] {$y\ind{\indn}(t)$};
    \draw (\startSpikex, \startSpikey) -- (\startSpikex + 2, \startSpikey) node[anchor = south west] {$\quad \spikes{\indn}$};
    \else
    \draw (\linestartX, \lineY) node[anchor = east] {$y\ind{I}(t)$};
    \draw (\startSpikex, \startSpikey) -- (\startSpikex + 2, \startSpikey) node[anchor = south west] {$\quad \spikes{I}$};
    \fi
}
    
\foreach \n in {-1,0,1}
    \tikzmath{
        \centerx = 1 + 1.8/2;
        \centery = 1 - 3*(0.9+0.2) + 0.5*0.9 +\n*(0.1+0.025);
    }
    \fill (\centerx, \centery) circle (0.025);
    

\tikzmath{
    \MTEMTotalHeight = 4*0.9 + 3*0.2;
}

\foreach \n in {0,1,2}
{
    \tikzmath{
        \lineStartX = 1 - 0.5 - 3 - 1.5;
    }
        \ifnum \n < 2
        \tikzmath{\lineStartY = 1 - \n*(0.9+0.2) - 0.2/2;}
        \else
        \tikzmath{\lineStartY = 1 - (\n+1)*(0.9+0.2) - 0.2/2;}
        \fi

    \foreach \m in {0,1,2,3}
    {
        \tikzmath{\lineEndX = 1 - 0.5 -1.5;}
        \ifnum \m < 3
        \tikzmath{\lineEndY = 1 - \m*(0.9+0.2) +0.9/2;}
        \else
        \tikzmath{\lineEndY = 1 - (\m+1)*(0.9+0.2) +0.9/2;}
        \fi
        \tikzmath{
        \aX = 2*\lineStartX/4 +\lineEndX/2 ;
        \aY = 2*\lineStartY/4 +\lineEndY/2;
        }
        
        \tikzmath{
            int \indn;
            \indn= \n +1;
            int \indm;
            \indm= \m +1;
        }
        \ifnum \n =0
            \ifnum \m <3
                \draw[blue, thick]  (\lineEndX, \lineEndY) -- (\lineStartX, \lineStartY) node[midway, above, font=\boldmath, sloped] {$a_{\indm,\indn}$};
            \else
                \ifnum \m = 3
                \draw[blue, thick]  (\lineEndX, \lineEndY) -- (\lineStartX, \lineStartY) node[midway, above, font=\boldmath, sloped] {$a_{I,\indn}$};
                \fi
            \fi
            
        \else
            \draw (\lineStartX, \lineStartY) -- (\lineEndX, \lineEndY);
        \fi
    }
    \tikzmath{
        int \indn;
        \indn= \n +1;
    }
    \ifnum \n < 2
    \draw (\lineStartX-1.5-1.1, \lineStartY) -- (\lineStartX-1.5, \lineStartY) node[anchor=west] {$x\ind{\indn}(t)$};
    \else
    \draw (\lineStartX-1.5-1.1, \lineStartY) -- (\lineStartX-1.5, \lineStartY) node[anchor=west] {$x\ind{J}(t)$};
    \fi
}

\end{tikzpicture}
\end{center}
    \caption{Sampling setup: $J$ input signals $x\ind{j}(t)$, $j=1\cdots J$ are mixed using a matrix $\mat{A}$ and produce signals $y\ind{i}(t)$, $i=1\cdots I$. Each $y\ind{i}(t)$ is then sampled using a time encoding machine TEM$\ind{i}$ which produces spike times $\spikes{i}$.}
    \label{fig: mixed TEM setup}
    \end{figure*}

In our setup, we let $\vect{y}(t)$ denote the vector signal composed of $y\ind{i}(t)$'s and let $\vect{y}(t)$ be such that
\begin{assumptions}
\item each $y\ind{i}(t)$ has a finite parametric representation:
\begin{equation}
y\ind{i}(t) = \sum_{k=1}^K c_{i,k}(\vect{y}) f_{k}(t),
\end{equation}
where the $c_{i,k}(\vect{y})$ are fixed coefficients that are unknown apriori and the $f_{k}(t)$'s, $k=1...K$ are known functions, \label{assume: parametric}
\item each $y\ind{i}(t)$ can be written as a linear combination of $x\ind{j}(t)$'s, $j = 1\cdots J$ where $J<I$:
\begin{equation}
\vect{y}(t) = \vect{Ax}(t),
\end{equation}
for a matrix $\vect{A} \in \mathbb{R}^{I\times J}$, and \label{assume: low-dim}
\item each $y\ind{i}(t)$ is sampled using a time encoding machine TEM$\ind{i}$ with parameters $\kappa\ind{i}$, $\delta\ind{i}$ and $\biasi{i}$ which are known and can vary between machines. The outputs of the machines are denoted $\spikes{i}$. \label{assume: time encoded}
\end{assumptions}

The sampling setup we described is depicted in Fig.~\ref{fig: mixed TEM setup}

We will consider two options for the functions $f_{k}(t)$:
\begin{subassumptions}
\item $f_{k}(t)$ is a sinc function
\begin{equation}
f_{k}(t) = \sinc_\Omega(t-\tau_k) = \frac{\sin \left(\Omega (t-\tau_k)\right)}{\pi(t-\tau_k)},
\end{equation}
for $\Omega$ and $\tau_k$ known, so that the $y\ind{i}(t)$'s are a finite sum of sincs, or \label{assume: sum of sincs}
\item $f_{k}(t)$ is a complex exponential function, 
\begin{equation}
f_{k}(t) = \exp \left(\mathbf{j}\frac{2\pi}{T}kt\right),
\end{equation}
so that the $y\ind{i}(t)$'s are bandlimited periodic functions. \label{assume: periodic}
\end{subassumptions}

Thankfully, the functions resemble each other enough for the treatment of the two functions to be done at the same time.
For both of them, we consider the reconstruction conditions with $\vect{A}$ satisfying either of the following two assumptions.
\begin{subassumptions}
\item The linear map from the low dimensional space $\vect{A} \in \mathbb{R}^{I\times J}$ is known. \label{assume: A known}
\item The linear map from the low dimensional space $\vect{A} \in \mathbb{R}^{I\times J}$ is unknown but the dimension of the low dimensional space $J$ is known. \label{assume: A unknown}
\end{subassumptions}

We first consider the case where $\vect{A}$ is known and provide conditions for perfect reconstruction in Section~\ref{sec: convergence proof} and a reconstruction algorithm in Sections~\ref{sec: rec alg 1}. We later provide applications for this scenario in Sections~\ref{sec: 1d space} and~\ref{sec: 2d space}, where we deal with time encoding video. 

Later, we will consider the case where $\vect{A}$ is unknown and provide a reconstruction algorithm based on singular value projection for low-rank matrix recovery in Section~\ref{sec: unknown A example}. We then follow with simulations to show results and with example applications for time encoding time-varying scenes.

\section{Known Low-Rank Factorization: Time Encoding and Reconstruction}
\label{sec: known low-dim map}

\subsection{Conditions for perfect reconstruction}
\label{sec: convergence proof}
We can establish the following sufficient conditions to ensure that a series of inputs $y\ind{i}(t)$ are reconstructible from their time encoding using machines TEM$\ind{i}$.

\begin{thm}
\label{thm: perfect reconstruction}
Let $I$ signals $y\ind{i}(t), i = 1\cdots I$ satisfy assumptions~\ref{assume: parametric},~\ref{assume: low-dim} and~\ref{assume: time encoded}, and their functionals $f_{k}(t)$ satisfy either of~\ref{assume: sum of sincs} or~\ref{assume: periodic} with the corresponding coefficients $c_{i,k}$ being drawn from a Lipschitz continous probability distribution. Now assume $\vect{A} \in \mathbb{R}^{I\times J}$ as defined in~\ref{assume: low-dim} is known and has every $J$ rows linearly independent. Then the inputs $y\ind{i}(t), i =1\cdots I$ are exactly determined by the spike times $\spikes{i}, i=1\cdots I$ if:
\begin{equation}
\label{eq: condition perf rec}
\sum_{i=1}^I \min \left(n_{\mathrm{spikes}^{(i)}}, K \right)  > JK,
\end{equation}
if the time encoding machines start sampling at $t_0$ with a known integrator value $\zeta_0\ind{i}$.
\end{thm}

The integrator value $\zeta_0\ind{i}$ indicates the value of the integral of TEM$\ind{i}$ at time $t_0$, before the time encoding begins.

An intuitive explanation of this result follows in Section~\ref{sec: Interpretation}.
The theorem just stated can be generalized to include scenarios where the initial integrator value is not known:
\begin{corollary}
\label{corol: perfect reconstruction - initial unknown}
Under the same assumptions of Theorem~\ref{thm: perfect reconstruction}, but when the time encoding machines have an unknown integrator value $\zeta_0\ind{i}$, the inputs $y\ind{i}(t), i =1\cdots I$ are exactly determined by the spike times $\spikes{i}, i=1\cdots I$ if:
\begin{equation}
\sum_{i=1}^I \min \left(n_{\mathrm{spikes}^{(i)}}-1, K \right)  > JK \label{eq: corol condition perf rec},
\end{equation}
\end{corollary}

We can prove the above theorem and corollary by writing it as a problem of rank one measurements, also called bi-linear measurements in~\cite{pacholska2020matrix}. The full proof is provided in Appendix~\ref{sec: appendix bilin mat meas}.

\subsection{Reconstruction Algorithm}
\label{sec: rec alg 1}
The spike time outputs of the machines $\spikes{i}$ provide constraints on the integral of the input signals:
\begin{equation}
\label{eq: signal integral btw spike times - in text}
\int_{t_\ell\ind{i}}^{t_{\ell+1}\ind{i}} y\ind{i}(u) \, du = 2\kappa\ind{i}\delta\ind{i} - \beta\ind{i}(t_{\ell+1}\ind{i} - t_{\ell}\ind{i}) =: b_\ell\ind{i}.
\end{equation}

These measurements can be rewritten to fit the rank one measurements formulation~\cite{pacholska2020matrix}.
Letting $\vect{C}(\vect{x})$ denote the matrix of coefficients $c_{j,k}(\vect{x})$ for the underlying signals $x\ind{j}(t)$, we can reconstruct $\vect{C}(\vect{x})$ (and therefore $\vect{y}(t)$) by solving
\begin{equation}
\label{eq: lin sys to solve}
b_\ell\ind{i} = \mathrm{vec}\left(\vect{a}_i \vect{F}(t_\ell\ind{i})^T \right) \mathrm{vec}\left(\vect{C}(\vect{x})\right),
\end{equation}
where $b_l\ind{i}$ is known and denotes the integral $\int_{t_0\ind{i}}^{t_\ell\ind{i}} y\ind{i}(u)\,du$ with $t_0\ind{i}$ denoting the time at which TEM $i$ starts integrating and
\begin{equation*}
    \left[\vect{F}(t_\ell\ind{i})\right]_k = \int_{t_0\ind{i}}^{t_\ell\ind{i}} f_k(u)\, du.
\end{equation*}



Under the conditions of Theorem~\ref{thm: perfect reconstruction}, the linear system in~\eqref{eq: lin sys to solve} is full rank and $\vect{C}(\vect{x})$ can be recovered perfectly.
Once the matrix $\vect{C}(\vect{x})$ has been recovered, one can recover the coefficients $c_{i,k}(\vect{y})$ of the $y\ind{i}(t)$'s by setting $\vect{C}(\vect{y}) = \vect{AC}(\vect{x})$ and can therefore recover the original sampled signals.

\subsection{Interpretation}
\label{sec: Interpretation}
The results in Theorem~\ref{thm: perfect reconstruction} and Corollary~\ref{corol: perfect reconstruction - initial unknown} establish a Nyquist-like criterion for recovery. They specify how to count the number of linearly independent constraints in the multi-channel TEM setup and require as many of these constraints as there are degrees of freedom to recover the sampled signals. The results can be summarized by a few key points:
\begin{enumerate}
\item When sampling a collection of signals with a known linear mapping to or from a lower dimensional representation, what matters is the number of degrees of freedom in the low dimensional space, rather than the number of degrees of freedom in the high dimensional space. More practically, to ensure perfect reconstruction, we need the number of linearly independent constraints to be at least the number of degrees of freedom in the low dimensional space $JK$. In the case where $J<<I$, we can see how this can be a major improvement in spiking rate.
\item When multiple correlated signals are sampled using different time encoding machines, a lower spiking rate of one machine can be compensated for by higher spiking rates from others. This can be seen by observing the summation in~\eqref{eq: condition perf rec} and noting that the \emph{total} spiking rate of the machines matters more than the individual spiking rates.
\item One machine can only compensate for another machine's low spiking rate up to a certain degree. This can be seen by the $\min$ term in~\eqref{eq: condition perf rec} which implies that every machine has a maximal ``useful'' spiking rate depending on the signal and that going above this spiking rate does not add further information.
\end{enumerate}
This has a series of implications. First, signals that have lower dimensional representations can be sampled at lower rates, increasing sampling efficiency. Second, if TEMs have limited capacity in terms of spiking rates (for example they have a refractory period), this can be compensated for by adding more TEMs. This would still ensure reconstruction of the input since the reconstruction condition in~\eqref{eq: condition perf rec} is only linked to the number of degrees of freedom in the low dimensional space. Third, we will see in Section~\ref{sec: 2d space} how the results help us solve time encoding of time-varying spatial signals which have certain structure in space.

Note that these results provide a stark improvement to sampling high-dimensional but low-complexity signals using regular clock-based sampling. In fact, Theorem~\ref{thm: perfect reconstruction} holds because of one key element: different dimensions of the signal are sampled at $\emph{different}$ times with continuous probability distributions. Regular-based sampling does not have this property; and indeed, it only takes a short mental exercice to see that the recovery of $\vect{y}(t)$ takes $IK$ samples if the $y\ind{i}(t)$'s are all sampled at the same sampling times.

To be fair, one \emph{could} ensure that different $y\ind{i}(t)$'s are sampled at different times (minus the continuous probability condition), but this condition is much more elegantly ensured in the time encoding scenario. Moreover, using different clocks in the classical sampling setup poses difficulties because it is hard to align different clocks. Clock alignment is not an issue in time encoding because the time reference of different time encoding machines can always be aligned by simply adding the spike trains of two machines and registering the time differences.

\section{Representing 2D Signals with Spikes: Videos with 1 Spatial Dimension}
\label{sec: 1d space}
The results obtained in previous sections provide considerable improvements in sample requirements for multi-signal reconstruction when these signals have a low dimensional structure. 

However, one cannot help but wonder how restrictive the conditions we have set are and which existing situations actually satisfy the given restrictions.

To answer these questions, we study how bandlimited videos fit into our framework.

First, we start with a simpler case and consider a two-dimensional (2D) signal $y(d,t)$ that is bandlimited in both components. Note that, exclusively when using term ``2D'' and ``3D'', when we refer to ``dimension'', we mean the spatial and time dimensions, i.e. the signal varies along each of these two components. We do not refer to the complexity of the signal (as it relates to $I$, $J$ and $K$) as we did before. 

\begin{flalign*}
    y(&d,t) \\
    &= \sum_{k_0=-K_0}^{K_0} \sum_{k_1=-K_1}^{K_1} c_{k_0,k_1}(y) \exp (j2\pi(\frac{tk_0}{T} +\frac{dk_1}{D}))  \\
    &=\sum_{k_0=-K_0}^{K_0} \sum_{k_1=-K_1}^{K_1} c_{k_0,k_1}(y) \\
    & \phantom{{}=\sum }\exp (j2\pi(\frac{tk_0}{T})) \exp ( j2\pi\frac{dk_1}{D}),
\end{flalign*}
where $c_{k_0,k_1}$ denote the 2D Fourier series coefficients of $y(d,t)$. Note that we assume that $y(d,t)$ has $(2K_0+1)\times (2K_1+1)$ of these coefficients with periods $T$ and $D$ in the time and space components, respectively.

\begin{figure}[tb]
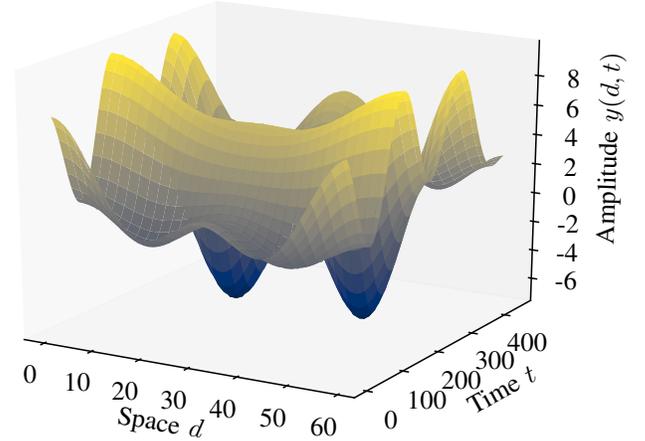

    \begin{minipage}[b]{0.85\linewidth}
        \centering
        \def\svgwidth{1.2\linewidth}

        \subfile{../figures/1d_space.tex}
    \end{minipage}
    \caption{2D signal $y(d,t)$ with spatial component $d$ and time component $t$.}
    \label{fig:1d space}
    \end{figure}

The results here will concern any such signal but, to make the treatment more intuitive, we will assume that we are dealing with a visual scene that has one continuous spatial component $d$ and is varying along time $t$. To be clearer, taking a picture of this scene at time $t$ provides the light intensity along one direction, which we assume to be the horizontal direction, without loss of generality. See Fig.~\ref{fig:1d space} for illustration.

Now assume that we sample this time-and-horizontally-varying scene using $I$ TEMs. Each TEM$\ind{i}$ is associated with a location in ``space", i.e. a position on the horizontal axis, $d\ind{i}$, such that the sampled signal $y\ind{i}(t)$ satisfies:

\begin{equation}
y\ind{i}(t) = y(d\ind{i}, t).
\end{equation}

To make the connection to the theory in Section~\ref{sec: known low-dim map}, we first need to define an auxiliary vector signal $\vect{x}(t)$ with $2K_1+1$ components, such that
\begin{align}
\label{eq: x in 2D periodic}
x\ind{k_1}(t) = \sum_{k_0=-K_0}^{K_0} c_{k_0, k_1}(\vect{y}) \exp (j\frac{2\pi}{T}k_0t).
\end{align}

We now notice that we can rewrite
\begin{align}
y\ind{i}(t) = y(d\ind{i},t) = \sum_{k_1=-K_1}^{K_1} x\ind{k_1}(t) \exp (j\frac{2\pi}{D}k_1d\ind{i}).
\end{align}

We can directly see that this brings back the structure we saw earlier, we have $\vect{y}(t) = \vect{A}\vect{x}(t)$ where $\vect{x}(t)$ is as defined in~\eqref{eq: x in 2D periodic} and $\left[\vect{A}\right]_{i,k_1} = \exp (-j\frac{2\pi}{D}k_1d\ind{i})$, where $i$ denotes the sampled channel.





\section{Representing 3D Signals with Spikes: Videos with 2 Spatial Dimensions}
\label{sec: 2d space}
\subsection{Theory}
We can use a similar treatment to understand how to time encode and reconstruct 3D signals $y(d_1, d_2, t)$. These signals can be interpreted as scenes that have 2 spatial components $d_1$ and $d_2$ (horizontal and vertical) and one time component $t$, as in videos.

We again assume that such a signal $y(d_1, d_2, t)$ is bandlimited along all components:
\begin{align}
y(d_1, d_2, t) = & \sum_{k_0=-K_0}^{K_0}\sum_{k_1=-K_1}^{K_1}\sum_{k_2=-K_2}^{K_2} c_{k_0,k_1,k_2}(y). \notag\\
&\exp \left(\mathbf{j}2\pi\left(\frac{tk_0}{T} +\frac{d_1k_1}{D_1} + \frac{d_2k_2}{D_2} \right)\right)
\label{eq: def 2d space}
\end{align}

\begin{figure*}[h!t]  
\begin{minipage}[b]{\linewidth}
        \centering
        \centerline{\includegraphics[width=0.8\columnwidth]{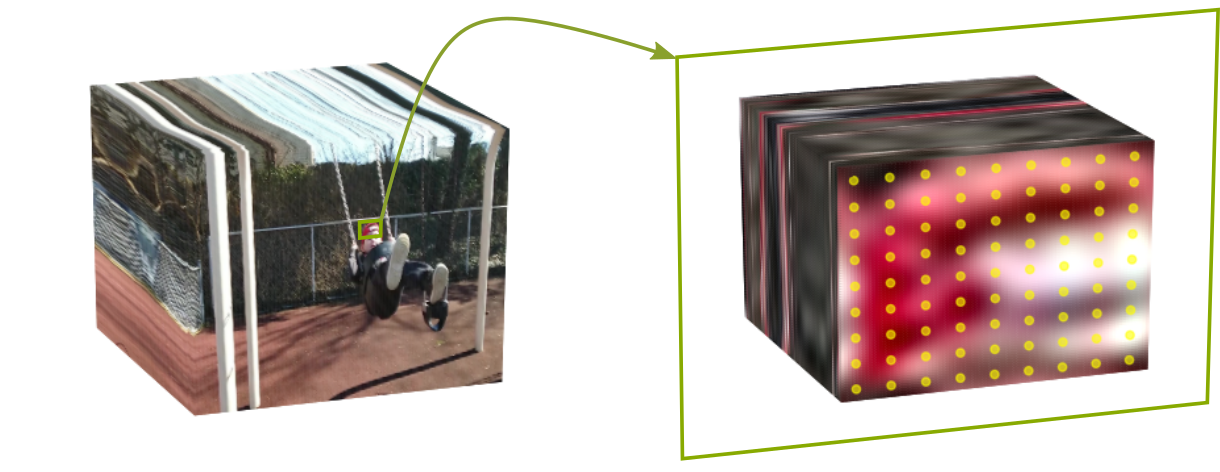}}

    \end{minipage}
    \caption{\label{fig: scene example} A time-varying scene is sampled at different spatial locations using TEMs. On the left, we see the scene with varying spatial and time components taken from the Need for Speed dataset~\cite{galoogahi2017need}. On the right, we see a time-varying patch which we will record using time encoding machines placed at the yellow dots. Originally, the video data we use was captured using a standard frame-based camera. We smoothly interpolate the video by assuming that the underlying structure is bandlimited and periodic and we aim to estimate the corresponding Fourier Series coefficients using the spikes emitted by the TEMs. In this case, we use a 9$\times$9 grid of TEMs on an interpolated version of a 9$\times$9$\times N_{f}$ video patch where $N_{f}$ is the number of frames used for the interpolation and time encoding. Therefore the number of Fourier Series coefficients to obtain is 9$\times$9$\times N_{f}$.}
\end{figure*}

Once again we assume that $y(d_1,d_2,t)$ is sampled in space at locations specified by $\vect{d}$ where sample $i$ is taken at spatial location $d\ind{i} = (d_1\ind{i}, d_2\ind{i})$ for some $d_1\ind{i}$ and $d_2\ind{i}$ in $\mathbb{R}$. An example is provided in Fig.~\ref{fig: scene example}. 

We define $x\ind{k_1, k_2}(t)$ in a similar fashion to~\eqref{eq: x in 2D periodic}: 

\begin{equation}
x\ind{k_1, k_2}(t) = \sum_{k_0=-K_0}^{K_0} c_{k_0,k_1,k_2}(y)\exp \left(\mathbf{j}2\pi\left(\frac{tk_0}{T}\right)\right), \label{eq: x_k definition}
\end{equation} 

 and we obtain the input signals to the TEMs
\begin{align}
y\ind{i}(t) = & \sum_{k_1=-K_1}^{K_1}\sum_{k_2=-K_2}^{K_2} x\ind{k_1,k_2}(t) \notag\\
& \exp \left(\mathbf{j}2\pi\left(\frac{d_1\ind{i}k_1}{D_1} + \frac{d_2\ind{i}k_2}{D_2} \right)\right).
\label{eq: 2d y=Ax}
\end{align}

Once more, we have found that we are time encoding $\vect{y}(t) = \vect{Ax}(t)$ with the entries of $\vect{x}(t)$ satisfying~\eqref{eq: x_k definition}, and a matrix $\vect{A}$ which is known if we know the locations of the time encoding machines $\vect{d}\ind{i}$. If the locations of the time encoding machines $\vect{d}\ind{i}$ are such that $\vect{A}$ has every $(2K_1+1)(2K_2+1)$ rows linearly independent, then all coefficients $c_{k_0, k_1, k_2}(t)$ can be recovered using $\prod_{n=0}^{2} (2K_n+1)$ appropriate measurements. This means that the continous scene can also be recovered, so we can interpolate the scene between spike times in both space and time components.

One example of a matrix $\vect{A}$ that satisfies the above constraint arises when one follows \emph{sufficient uniform gridding}.
\begin{definition}
\label{def: Nyq Sampl}
\emph{Sufficient Uniform Gridding} defines the sampling locations $d\ind{i}$ to follow a uniform grid over a spatial period, with $2K_1+1$ positions in the $d_1$ direction and $2K_2+1$ positions in the $d_2$ direction. More formally, $i$ ranges between zero and $(2K_1+1)(2K_2+1)$ and
\begin{align}
d_1\ind{i} &= \frac{\lfloor i/(2K_2+1) \rfloor}{2K_1+1} D_1,\\
d_2\ind{i} &= \frac{(i\mod (2K_2+1))}{2K_2+1} D_2.
\end{align}
\end{definition}

\begin{lemma}
\label{lemma: BL mix matrix satisfies} 
The matrix $\vect{A}$ obtained from using sufficient uniform gridding with entries as defined in~\eqref{eq: 2d y=Ax} has every $(2K_1+1)(2K_2+1)$ rows linearly independent.
\end{lemma}
\begin{proof}
The proof of relies on calculating the Gram matrix of $\vect{A}$ and noticing that it is diagonal and therefore full rank. Consequently $\vect{A}$ also has a full rank $(2K_1+1)(2K_2+1)$ and has every $(2K_1+1)(2K_2+1)$ rows linearly independent.
\end{proof}

This is not the only case in which $\vect{A}$ satisfies our assumptions, it seems that more general configurations of the spatial sampling can also work provided the samples cover the space.


As was the case in Section~\ref{sec: known low-dim map}, admitting that TEMs are receiving input signals that have a low dimensional structure allows one to manipulate the number of time encoding machines while keeping the same total spiking rate, and without compromising on reconstructibility.

In other words, every TEM does not have to be able to perfectly reconstruct its own input for the entire scene to be reconstructed. On the contrary, emitted spikes from all machines are used collaboratively in order to reconstruct the scene which has a parametric representation.

Therefore, if we have TEM-like receptors or sensors that have a limited spiking rate, spatial and temporal resolution can be regained by adding more sensors at new locations. 

\subsection{Simulations}

We would like to illustrate the theoretical results obtained in the previous section on an actual video, to illustrate the relationship between spatial and temporal sampling density.
First we choose a video recorded with a standard frame-based camera~\cite{galoogahi2017need}, and examine a patch of this video as shown in Fig.~\ref{fig: scene example}. This patch has  $H\times W \times N_f$ samples where $H$ refers to the height of the patch in pixels, $W$ refers to the width of the patch in pixels and $N_f$ refers to the number of frames. 

We assume the underlying scene has a periodic bandlimited structure (which is also the assumption that allows for finite uniform sampling). This allows us to (1) express the scene as in~\ref{eq: def 2d space} and (2) fix the corresponding number of Fourier series coefficients to match the number of samples $(2K_0+1)\times (2K_1+1) \times (2K_2+1)$ where $K_0 = \lfloor H/2\rfloor$ for example. The patch we consider is therefore a smooth function with a fixed number of parameters we are interested in and which can be sampled anywhere in time and space.

Given the smoothly varying patch, we place TEMs, for example, at the yellow dots in Fig.~\ref{fig: scene example}. 
In this case, we have a patch which is 9 pixels high and 9 pixels wide and we place a 9$\times$9 grid of time encoding machines, according to the definition of sufficient uniform gridding.
We will show, in our experiments that this is the minimum number of TEMs required to achieve perfect reconstruction.

We will also show how we can use more TEMs in the spatial components to obtain better resolution in the time component. This will not necessarily be the case the other way around: more sampling in time does not always provide improved spatial frequency resolution.

\begin{figure}
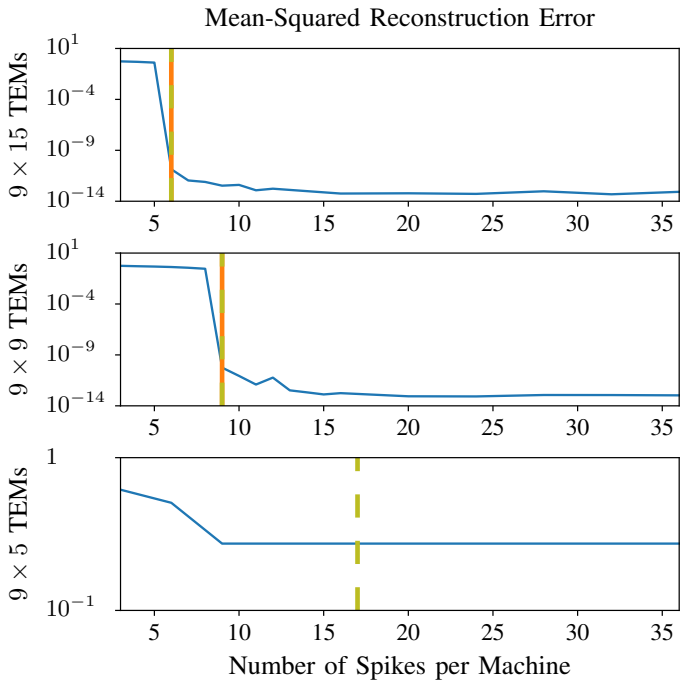

    \centering
    \def\svgwidth{\columnwidth}
    \subfile{../figures/n_spikes_sweep.tex}
    \caption{\label{fig: vary spike rate} Mean-squared reconstruction error as the number of spikes per time encoding machine varies. We assume the video has $9\times 9\times 9$ Fourier series coefficients that we wish to recover. The first row shows the evolution of the error as number of spikes increases for $9\times 15$ uniformly spaced TEMs . The second row shows the evolution of the error for $9\times 9$ uniformly spaced TEMs placed according to sufficient uniform gridding. The third row shows the evolution of the error for $9 \times 5$ uniformly spaced TEMs. For each plot, the dashed green lines mark the number of spikes per machine starting which we have more constraints than unknowns, not accounting for linear independence. The vertical orange line marks the threshold provided by Theorem~\ref{thm: perfect reconstruction} and sets the number of spikes per TEM starting which we have more \textit{linearly independent} constraints than unknowns.}
\end{figure}

The interpolated patch from Fig.~\ref{fig: scene example} is sampled using a fixed number of TEMs and we vary the number of spikes per TEM to see how this effects the reconstruction error in Fig.~\ref{fig: vary spike rate}.

We consider three scenarios from top to bottom: we have a $9\times15$ uniformly spaced grid of TEMs, a $9\times9$ uniformly spaced grid of TEMs (similar to uniform sufficient gridding), and a $9\times5$ uniformly spaced grid of TEMs.

We examine the evolution of the reconstruction error as the number of spikes per TEM increases. The number of constraints provide by the spikes (dashed green lines in the figure) will not always match the total number of spikes at which the reconstruction error significantly decreases. The results rather match the predictions of Theorem~\ref{thm: perfect reconstruction}. In fact, Theorem~\ref{thm: perfect reconstruction} cannot place any guarantees on reconstruction for the case where there are fewer TEMs than spatial components (i.e. when we have a $9\times5$ grid of TEMs). In fact, perfect reconstruction is never possible: the system will always be underdetermined because of too few sensors in the spatial domain.

\begin{figure}
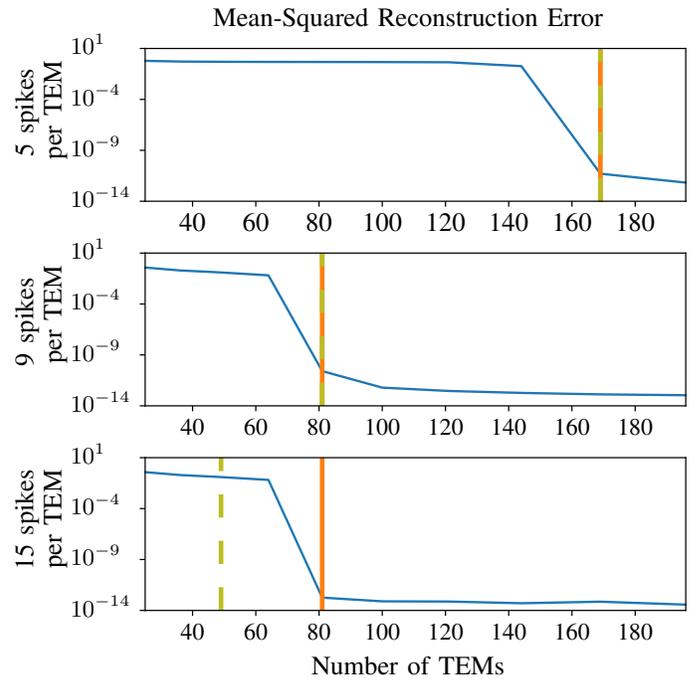

    \centering
    \def\svgwidth{\columnwidth}
    \subfile{../figures/n_tems_sweep.tex}

    \caption{\label{fig: vary num TEMs}
    Mean-squared reconstruction error as the number of spikes per time encoding machine varies. We assume the video has $9\times 9\times 9$ Fourier series coefficients that we wish to recover. The first row shows the evolution of the error as number of spikes increases for $5$ spikes emitted per machine. The second row shows the evolution of the error for $9$ spikes emitted per TEM which matches the sufficient rate starting which each TEM can perfectly reconstruct its input. The third row shows the evolution of the error for $15$ spikes per TEM.  For each plot, the dashed green line marks the number of TEMs starting which we have more constraints than unknowns, not accounting for linear independence. The vertical orange line instead marks the threshold provided by Theorem~\ref{thm: perfect reconstruction} and marks the number of TEMs starting which we have more \textit{linearly independent} constraints than unknowns.}
\end{figure}

On the other hand, we examine the scenario where we vary the number of TEMs for a fixed spiking rate per machine in Fig.~\ref{fig: vary num TEMs}. We similarly set the spiking rate to different levels from top to bottom: 5 spikes per TEM, 9 spikes per TEM and 15 spikes per TEM. Here, the sufficient number of spikes per machine $2K_0+1 = 9$  is the one that allows each machine to perfectly resolve its own input.

We notice that the reconstruction error undergoes a significant decrease once the number of TEMs is such that the condition of Theorem~\ref{thm: perfect reconstruction} is satisfied. As was the case for Fig.~\ref{fig: vary spike rate}, the threshold at which this decrease occurs does not depend on the total number of constraints (in green) but rather on the number of \textit{linearly independent} constraints.

We can draw a similar conclusion to that drawn for Fig.~\ref{fig: vary spike rate}: increasing the number of spikes per TEM beyond a certain point is not helpful and it is generally more beneficial to have more TEMs or sensors that spike less frequently.




\subsection{Coupling of Spatial and Temporal Resolution: Intuition and Consequences}
In a nutshell, the theory developed and experiments conducted all indicate that, if one would like to increase resolution, whether spatial or temporal, it is better to increase spatial sampling density. Increasing spatial sampling density is always useful, unlike increasing the number of spikes per machine.

In fact, a TEM can only output as much information as it receives, so if a TEM perfectly characterizes its own input using 15 spikes, there is no point in generating 20, 30 or 40 spikes.

On the other hand, increased spatial sampling can aid spatial \textit{and} temporal resolution because TEMs located at different locations will almost surely spike at different \emph{times} because they either have different inputs or different initial conditions~\cite{adam2020sampling}, or both.

This particular characteristic is not met by standard, frame-based video recordings where all pixels record information at the same time. Unfortunately, when all information is recorded at the same time, any information obtained from oversampling in the spatial domain is redundant rather than contributing to better resolution in the time domain as is the case when times are asynchronous.

In practice, this means two things: (1) TEMs or event-based sensors that have a limited spiking rate can be compensated for by simply having more sensors in space and (2) it is better to increase sampling capacity in the spatial domain when performing time encoding because this can improve \textit{both} spatial and temporal resolution.

\section{Unknown Low-Rank Factorization}
\label{sec: unknown A example}
\subsection{Problem Formulation and Algorithm}
We revisit the setup exposed in Section~\ref{sec: Problem Setup}. So far, we have assumed that we are given the time encodings of a collection of signals $y\ind{i}(t)$ with a low dimensional structure which we can reach by a known linear transformation $\vect{A} \in \mathbb{R}^{I\times J}$ and that we are asked to reconstruct the inputs $y\ind{i}(t)$. While this is a useful model in itself, we are also interested in studying the case where the linear transform $\vect{A}$ is unknown.

Once again, we assume we have the time encodings of a collection of signals $y\ind{i}(t)$ which satisfy assumptions~\ref{assume: parametric},~\ref{assume: low-dim} and~\ref{assume: time encoded}. Furthermore, we assume the functions $f_k(t)$ of $y\ind{i}(t)$ satisfy either of~\ref{assume: sum of sincs} or~\ref{assume: periodic} and that the linear transformation $\vect{A}$ is unknown as in~\ref{assume: A unknown}.

We wish to recover the signals $y\ind{i}(t), i = 1...I$ from their time encoding, with as few samples as possible.

To do so, we aim to reconstruct the coefficients of the parametric representation of $\vect{y}(t)$, $c_{k_0,k_1,k_2}$ as defined in~\ref{assume: parametric}. These coefficients are placed in the matrix $\vect{C}(\vect{y})$, with row $i$ containing the coefficients of signal $y\ind{i}(t)$. We note once more that $\vect{C}(\vect{y})$ can be written:
\begin{equation}
\vect{C}(\vect{y}) = \vect{AC}(\vect{x})\notag
\end{equation}
where $\vect{A} \in \mathbb{R}^{I \times J}$, $\vect{C}(\vect{x}) \in \mathbb{R}^{J \times K}$, $J<I$ and $J$ is known.

In words, $\coeffs{y}$ is a matrix which has a low rank matrix decomposition with a known rank.

The matrix $\coeffs{y}$ is probed using a sensing operator which we will call $\mathcal{S}$. The sensing operator performs the measurements in~\eqref{eq: lin sys to solve}, i.e. 
\begin{equation}
\label{eq: sensing operator definition}
\mathcal{S}_n\left( \coeffs{y}\right) = b_n, 
\end{equation}

where we index a pair $(i, \ell)$ by $n$.

Given this measurement setup, we can adopt the Singular Value Projection approach to recover the matrix $\coeffs{y}$ from few measurements~\cite{jain2010guaranteed}. 

The Singular Value Projection (SVP) algorithm alternately applies the low-rank constraint and the measurement constraint on the matrix of interest $\coeffs{y}$. In Algorithm~\ref{alg: SVP} we let $X^t$ be the estimate at iteration $t$ of the target matrix to reconstruct (in our case this is $\coeffs{y}$) and $Y^t$ be a proxy matrix to perform the iterations.

\begin{algorithm}
\caption{\label{alg: SVP}Singular Value Projection}
\begin{algorithmic}[1]
\REQUIRE $\mathcal{S}$, $b$, tolerance $\epsilon$, $\eta_t$ for $t = 0,1,2...$\\
\STATE{$X^0 = 0$ and $t=0$}
\REPEAT 
\STATE{$Y^{t+1}\leftarrow X^t - \eta_t\mathcal{S}^T(\mathcal{S}(X^t)-b)$}
\STATE{Compute top $J$ singular vectors of $Y^{t+1}: U_J, \Sigma_J, V_J$\\
\STATE$X^{t+1}\leftarrow U_J\Sigma_JV_J^T$}
\STATE{$t\leftarrow t+1$} 

\UNTIL{$\norm{\mathcal{S}(X^{t+1} - b)}_2^2 \leq \epsilon$}
\end{algorithmic}
\end{algorithm}

The SVP algorithm is based on projected gradient descent. Reconstruction guarantees for this algorithm were initially established in cases where the sensing operator satisfies the Restricted Isometry Property~\cite{jain2010guaranteed,candes2009exact,recht2010guaranteed}. This property does not hold in our case, given that our measurement operators have rank one. The rank one scenario has been treated in~\cite{zhong2015efficient} where Gaussianity assumptions are made on the measurement operators. Again, these assumptions do not hold for our case and we leave the theoretical analysis of convergence for future work. We do, however, illustrate the utility of our approach with simulations in  the next section. 

\subsection{Simulations}

\begin{figure}[tb]
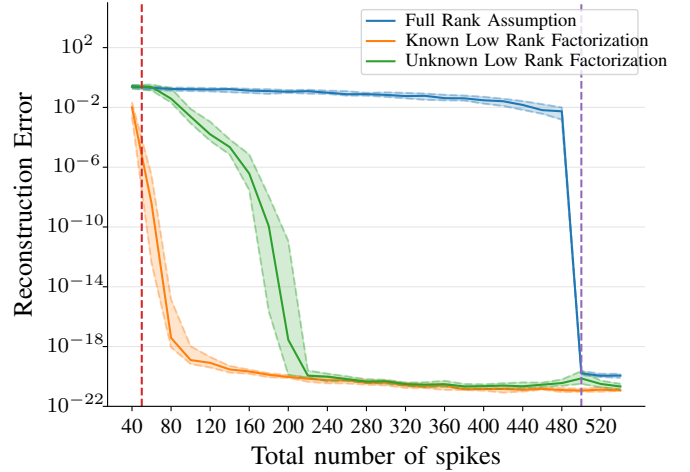
  
        \centering
        \def\svgwidth{0.95\columnwidth}
        \subfile{../figures/mixing_comparison.tex}

    \caption{\label{fig: reconstruction error} Reconstruction error of one out of twenty signals that have rank two, as the number of emitted spikes increases. The red dashed line marks the perfect reconstruction condition assuming the transform to the low dimensional space is known, and the purple dashed line marks the perfect reconstruction condition assuming there is no lower dimensional representation of the signals. We show the median and quartiles of the reconstruction error for 25 random trials, when assuming the signals have no low dimensional structure, when assuming they have a low dimensional structure with a known linear mapping, and when they have a low dimensional structure with an unknown linear mapping.}
\end{figure}

We provide simulation results to evaluate the reconstruction performance in different regimes.
We consider the scenario where we are time encoding and reconstructing twenty signals that are composed of 25 $\sinc$ functions at known locations and that can be written as linear combinations of two such signals.

We evaluate the reconstruction performance that varies with the number of spikes of all machines increase uniformly. We do this in the following cases:
\begin{itemize}
    \item[(S1)] when assuming the signals have no underlying low dimensional structure
    \item[(S2)] when assuming the signals have an underlying low dimensional representation which we can reach through a known linear transform $\vect{A}$, and
    \item[(S3)] when the signals have an underlying low dimensional representation with an unknown mapping $\vect{A}$.
\end{itemize}

For each of these cases, we time encode and reconstructing all twenty signals and compute the obtained normalized mean-squared error for the first signal among the twenty, assuming a random mapping to low dimensional space $\vect{A}$. Then we plot the median and quartiles of the mean-squared error on a log plot to compare performance. Results are included in Fig.~\ref{fig: reconstruction error}.

Note that, if we assume no underlying low dimensional structure (S1), the signals should be reconstructible assuming there are $I\times K$ linearly independent constraints. In this case, since the number of spikes of all machines increase uniformly, we will need $I\times K = 500$ spikes.
As for the scenario (S2), according to Theorem~\ref{thm: perfect reconstruction}, the signals should be reconstructible assuming that there are $J\times K$ linearly independent constraints. As before, this means we would need $J\times K = 50$ spikes.

We draw each of these conditions in Fig.~\ref{fig: reconstruction error} to see if the performance is consistent with our expectations.

Notice that assuming we know a transformation $\vect{A}$ to a low-dimensional space (S2) greatly improves reconstructibility compared to when we assume that there is no low rank structure for the input: the error decays much earlier in the first case than it does in the second case.

Assuming such a transformation exists but that we do not know it (S3), also offers benefit. While the reconstruction algorithm can be quite unstable in regimes where the number of spikes is not sufficient, it can yield a very good reconstruction for a higher number of spikes, where the scenario (S1) fails entirely.



\section{Conclusion}
\label{sec: Conclusion}

We have shown how time encoding can be used to encode and reconstruct multiple signals that have lower-dimensional representations.

The general case can be treated by reformulating our problem as a rank-one matrix measurement problem: we have shown that signals that have a known lower dimensional representation require fewer spikes for perfect reconstruction than if this lower dimensional representation did not exist.

Time encoding videos can then be rewritten as a special case of low-rank signal estimation. As a consequence, we show through theory and experiments that, if one wishes to increase spatial or temporal resolution, it is better to sample densely in space than to have TEMs emit more spikes. More practically, in the case of an event-based camera, it is better to have more pixels that fire asynchronously than to have pixels that fire more often.

Finally we have also examined the case where the signals of interest are low rank but we do not know the transformation to the low rank space. We applied low rank factorization algorithms and found significant experimental improvements compared to the case where no low rank structure is assumed.

In future work, we would like to further investigate low rank factorization within the time encoding setup and understand how it can be used to encode multi-dimensional data with a different structure to  that presented in the paper.

\IEEEpeerreviewmaketitle

\begin{appendices}

\section{Known Low Dimensional Mapping - Elaboration and Proof of Theorem~\ref{thm: perfect reconstruction}}
\label{sec: appendix bilin mat meas}

To prove Theorem~\ref{thm: perfect reconstruction}, we will use results about rank-one matrix measurements~\cite{pacholska2020matrix}.
The work in~\cite{pacholska2020matrix} assumes that one is attempting to reconstruct a matrix $\vect{C}$ using measurements of the form:
\begin{equation}
\label{eq: bilin meas nonvec}
b_n = \vect{g_n}^T \vect{C}\vect{h_n},
\end{equation}
and rewrites the measurements as 
\begin{equation}
\label{eq: bilin meas vec}
b_n = \mathrm{vec}(\vect{g_n}\vect{h_n}^T)^T \mathrm{vec}(\vect{C}).
\end{equation}

Note that we adopted a change of notation with respect to~\cite{pacholska2020matrix} to avoid confusion.

The results of~\cite{pacholska2020matrix} then hold under two further assumptions.
\begin{assumptions}
    \item $\vect{h_n}$ can be parametrized by one variable $t \in \mathbb{R}$. More precisely, we assume the $k$-th entry of $h_n$ has the form $[h_n]_k = h_k(t_n)$ where $h_k: \mathcal{I}\rightarrow \mathbb{R}, k = 0, \cdots, K-1$ are linearly independent functions from a linear space of fucntions $\mathcal{F}$, $\mathcal{I} \in \mathbb{R}$ is an interval or the whole real line and $t_n \in \mathcal{I}, n = 0, .. N-1$ are sampling times. Moreover, we assume that the sampling times $(t_0, ... t_{N-1})$ follow a continuous probability distribution on $\mathcal{I}^N$ and that for every non-zero element $h \in \mathcal{F}$, the set of zeros of $h$ has Lebesgue measure $(\lambda)$ equal to zero: $\lambda(\{t|f(t)=0\}) = 0$. \label{assume: f param}
    \item The vectors $\vect{g_n}$ are taken from a set $\mathcal{A}$, where every $J$ elements of $\mathcal{A}$ are linearly independent. \label{assume: g suff diff}
\end{assumptions}

As a result, a uniqueness condition can be obtained.
\begin{thm}[Pacholska '20]
\label{thm: bilin matrix measurements}
Consider the set of $KJ$ vectors of the form $vec(g_nh_n^T)$. It is a basis in $\mathbb{R}^{KJ}$ if and only if no more than $K$ vectors $g_n$ are equal.
\end{thm}

We are able to rewrite our problem as a rank-one measurement problem by letting each index $n$ denote a pair $(\ell, i)$, letting $b_n$ denote the integral $\int_{t_0}^{t_\ell\ind{i}}y\ind{i}(u)\, du$, letting $\vect{g_n}$ denote rows of the matrix $\vect{A}$ and letting $[\vect{h_n}]_k$ denote the integral $\int_{t_0}^{t\ind{i}_\ell} f_k(u)\, du$. 

We will use the following lemmas to prove Theorem~\ref{thm: perfect reconstruction}.
\begin{lemma}
Under the assumptions of Theorem~\ref{thm: perfect reconstruction}, the spike times $\spikes{i}, i=1\cdots I$ follow a continuous probability distribution. \label{lemma: spike times continuous} 
\end{lemma}
\begin{proof}
This closely follows the proof in~\cite{pacholska2020matrix}.
\end{proof}

\begin{lemma}
Let the $f_k(t)$'s be the functionals as defined in~\ref{assume: sum of sincs} or~\ref{assume: periodic} and define $F_k(t) = \int_{t_0}^t f_k(u)\, du$. Then the $F_k$'s are linearly independent functions from a linear space of functions $\mathcal{F}$ which is the space of bandlimited functions. Moreover, every non-zero element of $\mathcal{F}$ has a set of zeros with Lebesgue measure equal to zero.
\label{lemma: h_k satisfy constraint}
\end{lemma}
\begin{proof}
This follows by construction of the $f_k$'s which are linearly independent, leading to their integrals being linearly independent. The second part of the lemma follows from the properties of bandlimited functions.
\end{proof}

We can now prove Theorem~\ref{thm: perfect reconstruction} and Corollary~\ref{corol: perfect reconstruction - initial unknown}.

\begin{proof}[Proof of Theorem~\ref{thm: perfect reconstruction}]
We will assume that we operate under the assumptions set out in Theorem~\ref{thm: perfect reconstruction}.

We start by showing that different constraints imposed by the time encoding machines can be written as in~\eqref{eq: bilin meas nonvec}. In fact, two consecutive spike times $t_\ell\ind{i}$ and $t_{\ell+1}\ind{i}$ from a machine TEM$\ind{i}$ impose a constraint on the integral of the concerned signal:
\begin{equation}
\label{eq: signal integral btw spike times}
\int_{t_\ell\ind{i}}^{t_{\ell+1}\ind{i}} y\ind{i}(u) \, du = 2\kappa\ind{i}\delta\ind{i} - \beta\ind{i}(t_{\ell+1}\ind{i} - t_{\ell}\ind{i}).
\end{equation}

We define $Y\ind{i}(t) = \int_{t_0}^t y\ind{i}(u)\, du$ to be the integral of the signal $y\ind{i}(t)$ between $t_0$ and any later time $t$. Given~\eqref{eq: signal integral btw spike times}, we can compute $Y\ind{i}(t_\ell\ind{i})$ for any spike time $t_\ell\ind{i}$:
\begin{equation}
\label{eq: integral measurements}
Y\ind{i}(t_\ell\ind{i}) = \int_{t_0}^{t_{\ell}\ind{i}} y\ind{i}(u) \, du = 2\ell\kappa\ind{i}\delta\ind{i} - \beta\ind{i}(t_{\ell}\ind{i} - t_0\ind{i}).
\end{equation}

We define this quantity to be $b_\ell\ind{i} := Y\ind{i}(t_\ell\ind{i})$ and denote the function $F_k(t) = \int_{t_0}^t f_k(u)\, du$. We then rewrite the right-hand side of~\ref{eq: integral measurements} in terms of the parametrization of $y\ind{i}(u)$:
\begin{align}
b_\ell\ind{i} &= \int_{t_0}^{t_{\ell}\ind{i}} \sum_{k = 1}^{K} c_{i,k}(\vect{y})f_{k}(u) \, du \notag\\
&= \sum_{k = 1}^{K} c_{i,k}(\vect{y}) \int_{t_0}^{t_{\ell}\ind{i}} f_{k}(u) \, du \notag\\
&= \sum_{k = 1}^{K} c_{i,k}(\vect{y}) F_{k}(t_\ell\ind{i})  \notag\\
& = \left[\vect{F}(t_{\ell}\ind{i})\right]  \left[\vect{C}(\vect{y})\right]_i^T\label{eq: measurement eq}
\end{align}

Where we defined $\left[\vect{F}( t_{\ell}\ind{i})\right]$ to be the vector of integrals $F_k(t_\ell\ind{i})$ for $k=1...K$. We also defined $\vect{C}(\vect{y})$ to be the matrix of coefficients $c_{i,k}(\vect{y})$ as defined in~\ref{assume: parametric} and $\left[\vect{C}(\vect{y})\right]_i$ is the $i^{th}$ row containing the coefficients for signal $y\ind{i}(t)$.

We further rewrite $\vect{C}(\vect{y}) = \vect{A}\vect{C}(\vect{x})$ (from~\ref{assume: low-dim}) and obtain $\left[\vect{C}(\vect{y})\right]_i = \left[\vect{A}\right]_i \vect{C}(\vect{x})$. We thus obtain:
\begin{align}
b_\ell\ind{i} = \left[\vect{F}(t_{\ell}\ind{i})\right] (\left[\vect{A}\right]_i\vect{C}(\vect{x}))^T\notag\\
b_\ell\ind{i} = \left[\vect{F}(t_{\ell}\ind{i})\right] \vect{C}(\vect{x})^T\left[\vect{A}\right]_i^T\notag\\
b_\ell\ind{i} = \left[\vect{A}\right]_i \vect{C}(\vect{x}) \left[\vect{F}(t_{\ell}\ind{i})\right]^T\label{eq: rewriting into bilin form}
\end{align}
We can thus reindex the above equations: we let every $n$ correspond to a single pair $(\ell, i)$ and let $b_n = b_{\ell}\ind{i}$, $\vect{g_n} = \left[\vect{A}\right]_i$ and $\vect{h_n} = \left[\vect{F}(t_{\ell}\ind{i})\right]$.

We can now see that the vectors $\vect{h_n}$ can be parametrized by one variable $t\in\mathbb{R}$ using a set of function $h_k$ which satisfy assumption~\ref{assume: f param}, as stated by Lemma~\ref{lemma: h_k satisfy constraint}. Moreover, according to Lemma~\ref{lemma: spike times continuous}, the spike times follow a continuous probability distribution, as required in assumption~\ref{assume: f param}.

We can also see that the vectors $\vect{g_n}$ just defined satisfy assumption~\ref{assume: g suff diff} by construction since this is a condition in Theorem~\ref{thm: perfect reconstruction}.

Then, we note that under the conditions of Theorem~\ref{thm: perfect reconstruction}, one can extract $KJ$ constraints that satisfy the constraints of Theorem~\ref{thm: bilin matrix measurements}, thus ensuring perfect reconstruction of the matrix of parameters $\vect{C}(\vect{x})$.
\end{proof}

\begin{proof}[Proof of Corollary~\ref{corol: perfect reconstruction - initial unknown}]
Using similar notation used for the Proof of Theorem~\ref{thm: perfect reconstruction}, we note that the value $b_\ell\ind{i}$ is not known, when the initial integrator values $\zeta_0\ind{i}$ are not known, instead, we know the value of
\begin{equation}
\tilde{b}_\ell\ind{i} = b_\ell\ind{i} + \zeta_0\ind{i} = \left[\vect{A}\right]_i \vect{C}(\vect{x}) \left[\vect{F}(t_{\ell}\ind{i})\right]^T + \zeta_0\ind{i}.
\end{equation}
Continuing in the same logic as before, we let $\tilde{b}_n = \tilde{b}_{\ell}\ind{i}$, $\vect{g_n} = \left[\vect{A}\right]_i$ and $\vect{h_n} = \left[\vect{F}(t_{\ell}\ind{i})\right]$.

We then obtain
\begin{equation}
\tilde{b}_n = \mathrm{vec}(\vect{g_nh_n}^T)^T\mathrm{vec}(\vect{C(x)}) + \zeta_0\ind{i_n}.
\end{equation}

To keep things in matrix form, we first denote $\mathbf{G_n} = \mathrm{vec}(\vect{g_nh_n}^T)^T$ and then denote the row vector $\mathbf{\tilde{G}_n} = [\mathbf{G_n}, e_{i_n}]$ where $e_{i_n}$ is a length-$J$ row vector with a 1 in the column corresponding to the machine that generated measurement $n$ and zeros otherwise. We also denote the column vector $\mathbf{\tilde{C}(x)} = [\mathbf{vec}(\vect{C(x)})^T, \zeta_0\ind{1}, \zeta_0\ind{2}, \cdots, \zeta_0\ind{J}]^T$.

The measurement therefore satisfies
\begin{equation}
\tilde{b}_n = \vect{\tilde{G}_n}\vect{\tilde{C}(x)}.
\end{equation}

For the system to be invertible we need $J\times(K+1)$ of the vectors $\vect{\tilde{G}_n}$ to be linearly independent.

If we satisfy the condition set by Corollary~\ref{corol: perfect reconstruction - initial unknown} in~\eqref{eq: corol condition perf rec}, we also satisfy the condition set by Theorem~\ref{thm: perfect reconstruction} in~\eqref{eq: condition perf rec}. This means that there are $JK$ rows $\vect{G_n}$ of $\vect{G}$ that are linearly independent.
Now let us consider the extension $\vect{\tilde{G}}$, the corresponding $JK$ rows from $\vect{G}$ will still be linearly independent (otherwise we reach a contradiction).

According to the assumptions of the corollary, $\vect{\tilde{G}}$ has a $K+1^{\textrm{st}}$ set of $J$ rows, each row coming from a TEM$\ind{j}$. Let $g_j$ be the $K+1^{\textrm{st}}$ row coming from TEM$\ind{j}$. The right part of the vector will be comprised of $e_j$ as is the case for all the other vectors coming from machine TEM$\ind{j}$. Therefore, this row is linearly independent of all rows coming from all other machines. 

The only remaining question is whether $g_j$ is independent of the first $K$ rows coming from the same TEM$\ind{j}$. Because the right part $e_j$ of the vectors are all the same, the $K+1^\textrm{st}$ element can only be a linear combination of the other $K$ elements if the linear combination coefficients all sum to 1. The latter case does not occur, with probability 1, thus showing that every $K+1^\textrm{st}$ constraint of a TEM is linearly independent of the other $K$ constraints of the machine, and showing that one can find $J\times(K+1)$ in $\tilde{G}$ that are independent, thus concluding our proof.
\end{proof}



\end{appendices}

\bibliographystyle{IEEEtran}
\bibliography{main}

\end{document}